# A Be Star + He Star Binary as an Indicator of a Binary Mass Transfer Phase

Yuchen Bao[1,2], Zhenwei Li[1,3], Hongwei Ge[1,3], Xuefei Chen[1,2,3], and Zhanwen Han[1,2,3]
[1] International Centre of Supernovae (ICESUN), Yunnan Key Laboratory of Supernova Research, Yunnan Observatories, Chinese Academy of Sciences (CAS),
Kunming 650216, People's Republic of China; lizw@ynao.ac.cn, cxf@ynao.ac.cn
[2] University of Chinese Academy of Sciences, Beijing 100049, People's Republic of China
[3] Key Laboratory for the Structure and Evolution of Celestial Objects, Chinese Academy of Science, People's Republic of China

## Abstract

The rapid rotation of Be stars is supposed to mainly originate from binary evolution. In recent years, more and more Be stars with helium (He) star companions have been discovered, which provides a significant opportunity to study binary interaction physics. In this work, we perform binary population synthesis with an updated binary mass transfer stability criterion and try to understand the details of mass transfer processes by constructing a series of Be star + He star (BeHe) binary populations. We found that the simulations and the observations can be divided into two groups according to the masses of components, corresponding to the two distinct evolutionary processes during the mass transfer. In particular, we found that the mass ratios of BeHe binaries may be taken as a probe of the initial mass ratios of the primordial binaries. Moreover, the results suggest that a higher mass transfer efficiency ($\gtrsim 0.5$) supports the observations better. The simulations predicted too many Be star binaries experiencing case B mass transfer, which conflicts with the observations. The reason is due to either observational selection effects or unclear physical factors.

*Unified Astronomy Thesaurus concepts:* Stellar evolution (1599); Mass ratio (1012); Binary stars (154); Be stars (142)

## 1. Introduction

More than half of stars are discovered in binary systems, and binary stars are the cornerstone of modern astrophysics (e.g., H. Sana et al. 2012, 2013; M. Moe & R. Di Stefano 2017; Z.-W. Han et al. 2020; Y. Guo et al. 2022; X. Chen et al. 2024 and references therein). Binary interaction is responsible for the formation of many special celestial objects, such as Type Ia supernovae, gravitational-wave (GW) sources, and X-ray binaries (e.g., Z.-W. Han et al. 2020; Z. Liu & R. J. Stancliffe 2020; D. Belloni & M. R. Schreiber 2023; X. Chen et al. 2024; Z. Li et al. 2024; A. J. Ruiter & I. R. Seitenzahl 2025). Though much progress has been made during the last two decades, there are still some questions in binary evolution that remain unsolved, i.e., the common envelope (CE) evolution, binary mass transfer stability, and mass transfer efficiencies (N. Ivanova et al. 2013b, 2020; X. Chen et al. 2024). In this paper, we proposed that Be star binaries can be taken as an important indicator for the binary mass transfer processes and would put constraints on the issues of binary mass transfer stability and the mass transfer efficiencies.

Be stars are a type of rapidly rotating main-sequence (MS) star with H$\alpha$ emission lines in their spectra (e.g., O. R. Pols et al. 1991; J. Zorec & D. Briot 1997; J. M. Porter & T. Rivinius 2003; M. V. McSwain & D. R. Gies 2005; T. Rivinius et al. 2013; T. Rivinius & R. Klement 2024). The emission features are generally thought to originate from the circumstellar disk surrounding the central star (e.g., O. Struve 1931; R. W. Hanuschik 1996; J. M. Porter & T. Rivinius 2003; A. C. Carciofi & J. E. Bjorkman 2006; A. C. Carciofi et al. 2009; T. A. A. Sigut et al. 2009; A. C. Carciofi 2011; A. Granada et al. 2013; T. Rivinius et al. 2013; R. O. Brown et al. 2019). The physical mechanisms responsible for the rapid rotation of Be stars are still under debate. The key to forming a star with high spin velocity is the transfer of angular momentum. One possible explanation is the single scenario; i.e., Be stars may spin up due to the gain of angular momentum from the molecular cloud at their birth or the transportation of angular momentum from the central contracting core to the outer envelope (G. Meynet & A. Maeder 2000; B. Mathew et al. 2008; B. Hastings et al. 2020; P. Kervella et al. 2022). Another commonly adopted point of view is the binary evolutionary scenario; i.e., the rapid rotation of Be stars is caused by the binary interaction (e.g., S. Kriz & P. Harmanec 1975; W. Packet 1981; O. R. Pols et al. 1991; S. E. de Mink et al. 2013, 2014; T. Rivinius et al. 2013; Y. Shao & X.-D. Li 2014; D. Boubert & N. W. Evans 2018; N. Langer et al. 2020; Y. Shao & X. D. Li 2020; B. Hastings et al. 2021). In binary evolutionary theory, the donor will transfer material to the accretor via the inner Lagrangian point as the star fills its Roche lobe, namely, stable Roche lobe overflow (RLOF). The accretor can increase its mass and the spin angular momentum during the mass transfer processes. Once the accretor is spun up to near-critical rotation, a classical Be star is formed (e.g., J. M. Porter & T. Rivinius 2003; R. H. D. Townsend et al. 2004; S. Ekström et al. 2008; W. Huang et al. 2010; T. Rivinius et al. 2013; J. Zorec et al. 2016).

Be stars produced via binary interaction typically have an envelope-stripped companion star or a compact object. In the observations, Be stars show several types of companion stars, such as neutron stars (NSs), white dwarfs (WDs), and helium (He) stars. Be star binaries with NS or WD companions can also present X-ray characteristics and are also known as Be X-ray binaries (BeXBs; e.g., L. Maraschi et al. 1976; S. Rappaport & E. P. J. van den Heuvel 1982; I. Negueruela 1998; N. V. Raguzova & S. B. Popov 2005; P. Reig 2011; D. Belloni & M. R. Schreiber 2023; D. R. Gies et al. 2023; C.-H. Zhu et al. 2023; H. Ge et al. 2024; K. A. Rocha et al. 2024; A. Marino et al. 2025). BeXBs may end their lives as double NSs or NS + black hole binaries, double black holes. Some of them may be identified as GW sources (e.g., M. Grudzinska et al. 2015; I. Ablimit & K. Maeda 2018; Y. Shao & X. D. Li 2018a, 2018b; N. Langer et al. 2020; D. Chattopadhyay et al. 2021). Be star + He star (BeHe) binaries are predicted to be abundant in the Milky Way. However, due to the complicated spectral features, only a dozen BeHe binaries are classified in the observations (e.g., J. Casares et al. 2014; L. Wang et al. 2017, 2018, 2021, 2022, 2023; R. Klement et al. 2022a, 2022b, 2024, 2025; Y. Nazé et al. 2022; S. Janssens et al. 2023; T. Rivinius et al. 2024).

BeHe binaries only experience one mass transfer phase, and they reserve important information about the progenitor binaries and the details of mass transfer. Therefore, BeHe binaries are an ideal laboratory for studying binary interaction physics (L. Yungelson et al. 2024; B. Hovis–Afflerbach et al. 2025). For example, BeHe binaries are formed through stable RLOF, which can indicate the critical condition between the dynamically unstable mass transfer and stable RLOF. Besides, the accurate estimation of the masses of the Be star and companion star would strongly constrain the mass transfer efficiencies. In recent works, L. Wang et al. (2021, 2023) conducted a long-term optical spectroscopic monitoring program for the detection of Be stars, where the radial velocities and the orbital period and physical parameters can be well determined. The results provide an important opportunity to study the formation of Be stars as well as the binary evolutionary theory.

In this work, we aim to understand the details of mass transfer processes by constructing a series of BeHe binary populations. We adopt the mass transfer stability criterion recently proposed by H. Ge et al. (2010, 2015, 2020a, 2024) in the binary population synthesis (BPS) simulations. This criterion has been used to simulate the formation of double WDs, and the observations can be well reproduced upon this new criterion (Z. Li et al. 2023). The remainder of this paper is organized as follows. In Section 2, we introduce the methods and input parameters used in the simulations. The main results, including the implications for the binary mass transfer stability and mass transfer efficiencies, are presented in Section 3. We offer our conclusions in Section 4.

## 2. Method and Input Parameters

In this work, we perform BPS via the rapid population synthesis code BSE (J. R. Hurley et al. 2000, 2002; P. D. Kiel & J. R. Hurley 2006), where important upgrades, including stellar wind and remnant-mass prescriptions, have been made (S. Banerjee et al. 2020). The single-star models are calculated based on the analytic formulae that approximate the stellar evolutions with a wide range of stellar mass and metallicity. Important physical processes during the binary interaction, such as GW radiation, tidal friction, stable RLOF, and wind accretion, have been considered. This study considers initial stellar masses ranging from 1 to 50 $M_\odot$, with a fixed metallicity of 0.02. The key input parameters are described below.

For a primordial binary system, the massive one may fill its Roche lobe at a certain evolutionary stage and initiate the mass transfer. If the mass transfer is dynamically unstable, the donor's envelope will engulf the companion star, and the binary enters into the CE phase (e.g., M. Livio & N. Soker 1988; R. F. Webbink 2008; N. Ivanova et al. 2013a, 2013b; K. A. Postnov & L. R. Yungelson 2014). If the mass transfer is dynamically stable, the envelope of the donor is removed via the stable RLOF. In the latter case, the accretor can gain angular momentum from the transferred material and may be spun up as a Be star. An important point is whether the mass transfer is stable or not, i.e., the mass transfer stability. The mass transfer stability can be determined by the so-called critical mass ratio, $q_c$ (massive one/less massive one); i.e., if the mass ratio at the onset of mass transfer is higher than $q_c$, the mass transfer is dynamically unstable, and the CE forms soon after. In this work, we adopt the results from H. Ge et al. (2020b), where the critical mass ratios are derived from realistic stellar structures based on the adiabatic mass-loss model. In their model, $q_c$ varies with the evolutionary stage of the donor (H. Ge et al. 2010, 2015, 2020a, 2024). Typically, for donors on the MS or in the Hertzsprung gap (HG), $q_c$ is in the range of 3–5, and for a donor initiating mass transfer at the giant branch, H. Ge et al. (2020a) give $q_c$ in a quite large range (1.2–10). H. Ge et al. (2020b) discussed that thermal timescale mass transfer through the other Lagrangian points may dominate the formation of the CE before triggering the dynamical timescale mass transfer. Therefore, we set the upper limit of $q_c$ to be 5.

For binaries experiencing stable RLOF, the mass transfer efficiency is one of the most uncertain parameters. The mass transfer efficiency, $\beta_{\rm rlof}$, is given by (J. R. Hurley et al. 2002)

$$\beta_{\rm rlof} \equiv \frac{\dot{M}_2}{|\dot{M}_1|} = \min\left(10\frac{\dot{M}_{\rm KH,2}}{\dot{M}_1}, \beta\right), \quad (1)$$

where $\dot{M}_1(<0)$ is the donor's mass-loss rate, $\dot{M}_2(>0)$ is the accretion rate, $\dot{M}_{\rm KH,2}$ is the thermal timescale mass transfer rate of the accretor, and $\beta$ is a free parameter that determines the specific value of the mass transfer efficiency. In this work, we consider four models with different inputs of $\beta$.

1. *Model I.* The $\beta$ value varies according to the accretor's spin. The mass transfer efficiency is reduced by a factor of $(1 - \Omega/\Omega_{\rm cr})$, where $\Omega$ and $\Omega_{\rm cr}$ are the angular velocity and the critical angular velocity of the accretor, respectively. If $\Omega$ is larger than $\Omega_{\rm cr}$, the transferred masses cannot be accreted, and the unprocessed material is assumed to be lost from the accretor as isotropic wind, carrying away the specific orbital angular momentum of the accretor, $j_{\rm iso} = (M_1/M_2)(J_{\rm orb}/M_T)$ (R. J. Stancliffe & J. J. Eldridge 2009). $M_1$ is the donor mass, $M_2$ is the accretor mass, $M_T \equiv M_1 + M_2$, and $J_{\rm orb}$ is the orbital angular momentum.
2. *Model II.* The $\beta$ value is kept constant at 0.1.
3. *Model III.* The $\beta$ value is kept constant at 0.5.
4. *Model IV.* The $\beta$ value is kept constant at 1.

The lost material is assumed to carry away the accretor's specific angular momentum. The angular momentum loss due to mass loss can be expressed as

$$\dot{J}_{\rm ML} = -(1 - \beta_{\rm rlof})\dot{M}_1\left(\frac{M_1}{M_1 + M_2}\right)^2\frac{2\pi a^2}{P_{\rm orb}}, \quad (2)$$

where $a$ is the binary separation and $P_{\rm orb}$ is the orbital period.

We generated $10^7$ primordial binaries for the Monte Carlo simulations. The following are the input parameters for BPS.

The initial mass of the primary star, $M_{1,i}$, follows the initial mass function (IMF) given by G. E. Miller & J. M. Scalo (1979) and P. P. Eggleton et al. (1989),

$$M_{1,i} = \frac{0.19X}{(1-X)^{0.75} + 0.032(1-X)^{0.25}}, \quad (3)$$

where $X$ is a random number and is limited between 0.4 and 0.999992, which approximately gives mass ranging from 0.1 to $100\,M_\odot$. In this work, only primary stars with masses in the range of $1$–$50\,M_\odot$ are evolved.

The initial mass of the secondary is determined by the initial mass ratio ($q_i$) distribution (T. Mazeh et al. 1992) as follows:

$$n(1/q_i) = 1, \quad q_i > 1, \quad (4)$$

where $q_i$ is the mass ratio of the primordial binary ($M_{1,i}/M_{2,i}$, where $M_{2,i}$ is the initial mass of the secondary and $M_{1,i} > M_{2,i}$).

The initial binary separation follows the distribution given by Z. Han (1998):

$$an(a) = \begin{cases} \alpha_{\rm sep}(a/a_0)^m, & a \leqslant a_0, \\ \alpha_{\rm sep}, & a_0 \leqslant a \leqslant a_1, \end{cases} \quad (5)$$

where $\alpha_{\rm sep} \approx 0.07$, $a_0 = 10\,R_\odot$, $a_1 = 5.75 \times 10^6\,R_\odot = 0.13$ pc, and $m = 1.2$. The metallicity is set to be 0.02, and the star formation rate is considered to be constant at $S = 3\,M_\odot\,{\rm yr}^{-1}$ (L. F. Smith et al. 1978; R. Diehl et al. 2006; T. P. Robitaille & B. A. Whitney 2010) over the past 15 Gyr.

## 3. Results

### 3.1. Be Star Binaries as a Probe of the Initial Binary Mass Ratio

A BeHe binary only experiences one mass transfer phase and would still retain important information on the primordial binary. In the first step, we try to understand the detailed formation history of the observational BeHe binaries. In the left column of Figure 1, we present the simulations of the four models and the observed samples in the $M_{\rm Be}$–$M_{\rm He}$ plane, where the initial mass ratios of the primordial binaries are shown in colored dots. The observed BeHe binaries taken from L. Wang et al. (2023) and R. Klement et al. (2025) are also shown for comparison. In model I, the mass transfer efficiency is determined by the specific mass transfer process and varies with a large range. We find that the averaged mass transfer efficiencies (averaged values of $\beta_{\rm rlof}$ in Equation (1)) decrease from 0.80 to 0.07 with increasing binary orbit periods, similar to the results of Y. Shao & X. D. Li (2016). In models II, III, and IV, the accretion is limited by the thermal timescale of the accretor, and the mass transfer efficiencies vary in a narrow range. In particular, for model IV, the mass transfer is near conservative with $\beta_{\rm rlof}$ larger than $\sim$0.95 (see also Y. Shao & X. D. Li 2021).

We first divide the simulated results into two groups, namely, group 1 and group 2, since we find that group 1 and group 2 show distinctive evolutionary pathways to Be star binaries. The two groups are shown in the middle and right columns in Figure 1, respectively. In group 1, the primordial binary experiences a mass transfer phase when the initial primary star (He star progenitor) evolves at the early-MS stage. Here the "early-MS stage" means that the primary star still does not develop an He core after 20% of the mass has been lost. Otherwise, we define it as the "late-MS stage." Mass transfer happens for donors at the late-MS stage or a more evolved stage (including the HG, red giant branch (RGB), and He core-burning stages) is classified as group 2. It should be noted that our classification of the mass transfer is different from the traditional manner, where the mass transfer is usually classified as case A (RLOF happens for the donor at the hydrogen-burning stage), case B (RLOF happens for the donor at the hydrogen shell-burning stage), and case C (RLOF happens for the donor after exhaustion of He in the core; K. A. Postnov & L. R. Yungelson 2014).

In group 2, we first find that the produced He star has a strong correlation with the initial primary star mass; i.e., massive He star progenitors generally produce massive He stars (see also L. Yungelson et al. 2024). BeHe binaries in group 2 can be divided into two cases. In the first case, the primary star fills its Roche lobe at the late-MS stage, and the He core is produced soon after the primary star loses a small part of its envelope mass. In the second case, the He core is produced when the primary star fills its Roche lobe at a more evolved stage (beyond the MS). For both cases, the final He star mass can be approximately determined by the initial primary star mass, since the He core would not increase a lot in the subsequent mass transfer phase (increase by no more than 10% of the He core mass at birth). However, in group 1, the primary star will lose a significant part of its mass during the MS stage, where the He core is still not produced. Therefore, the final He star mass is almost independent of the initial primary star.

Based on the above relation between the He star mass and its progenitor mass, we see that an approximate trend exists in group 2 for the four models; i.e., the large initial mass ratio generally leads to a small mass ratio ($M_{\rm Be}/M_{\rm He}$) of Be star binaries. The reason can be understood as follows. In model I, though the mass transfer efficiency for binaries in group 2 varies in the range of 0.04–0.45, the mass transfer for more than 95% of the binary systems is very inefficient (with $\beta \lesssim 0.1$). Then, the mass transfer efficiency can be approximately considered a constant in group 2 for all four models. For a binary with a fixed mass transfer efficiency, the binary with a massive secondary would lead to a massive Be star if the initial primary mass is equal. Therefore, the primordial binaries with large mass ratios generally lead to small mass ratios of Be star binaries. Such a relation is not found in group 1, since the final He star mass has a weak relation with the initial primary star mass, as explained above. We also find that the result in group 1 of model I has large dispersions, because the mass transfer efficiency for binaries in group 1 can vary in a large range (from 0.07 to 0.80).

The results in group 2 of models I and II are quite similar, as both cases have very inefficient mass transfer processes. The main difference is that there is dispersion for Be star binaries with $q \sim 7$ in model I, which are produced with a large mass transfer efficiency. With the increase of mass transfer efficiency, Be star binaries with larger mass ratios can be produced. In group 2, we see that most Be star binaries have mass ratios smaller than 9 and 12 for model III and model IV, respectively. In group 1, the typical mass ratios are 5–18, 4–12, 5–16, and 9–22 for models I, II, III, and IV, respectively.

Most of the observed Be star binaries have mass ratios in the range of 4–22 ($M_{\rm Be}/M_{\rm He}$), which can be covered by the results of models I, III, and IV. Model II predicts that no binaries with mass ratios greater than 12 can be produced, which is definitely not supported by the observations. Therefore, the preliminary results may suggest that inefficient mass transfer cannot explain the formation of Be star binaries. The detailed comparison of the simulated results and the observations is addressed in the following sections.

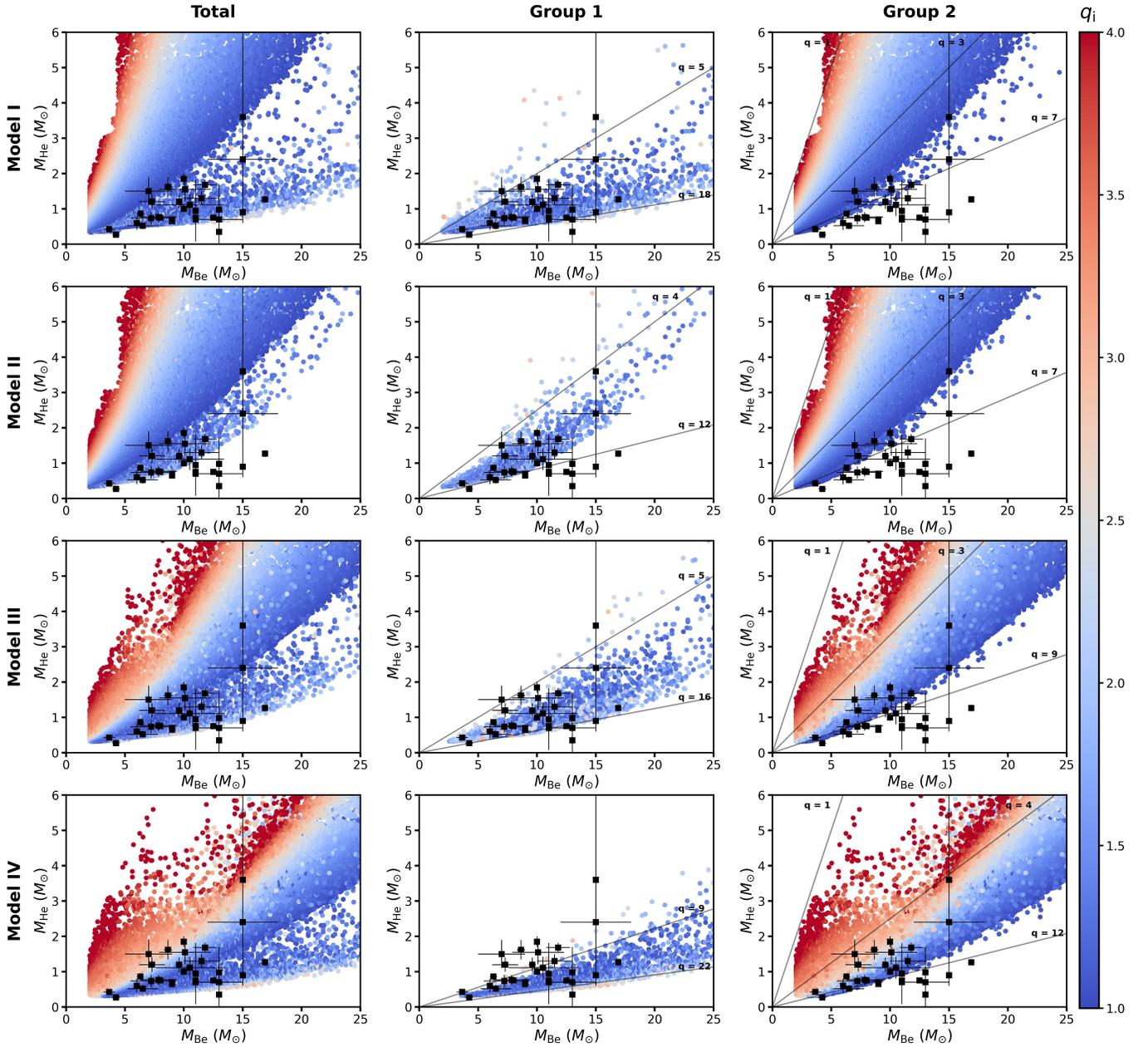

**Figure 1.** The $M_{Be}$–$M_{He}$ diagram for the four models, where the colored dots are for the initial mass ratios ($q_i$) of the primordial binaries. $q_i$ is defined as the ratio of the massive star to the less massive one in the primordial binary, i.e., $q_i = M_{1,i}/M_{2,i}$. The black squares represent the observed BeHe binaries reported by L. Wang et al. (2023) and R. Klement et al. (2025). All simulated Be star binaries are presented in the left panels. The simulations are divided into two groups based on different Be star evolutionary paths, as shown in the middle (group 1) and right (group 2) panels. In group 1, the mass transfer happens when the primary star is at the early-MS stage. In group 2, the mass transfer starts at a more evolved stage (see the main text for more details). The thin black lines correspond to mass ratios $q = M_{Be}/M_{He}$ equal to several constant values.

### 3.2. Population Properties of Be Stars

We then perform BPS to obtain the populations of BeHe binaries in the Galaxy. In our simulations, we find that the total number of Be star binaries in our Galaxy for the four models is $4.2 \times 10^5$, $5.6 \times 10^5$, $11.0 \times 10^5$, and $5.9 \times 10^5$, respectively.

It should be noted that there are more BeHe binaries in model III. The reasons can be understood as follows. First, in comparison to model I and model II, model III typically has large mass transfer efficiencies. This means that the primary star of the progenitor binary with a lower mass can produce a similar Be star in model III. According to the IMF, the number of low-mass stars is higher. Therefore, more Be star binaries are produced in model III. Second, in comparison to model IV, the lower mass transfer efficiencies in model III would lead to more Be stars with lower masses. In consideration of the longer lifetimes for low-mass Be stars, there are more Be star binaries in model III than in model IV.

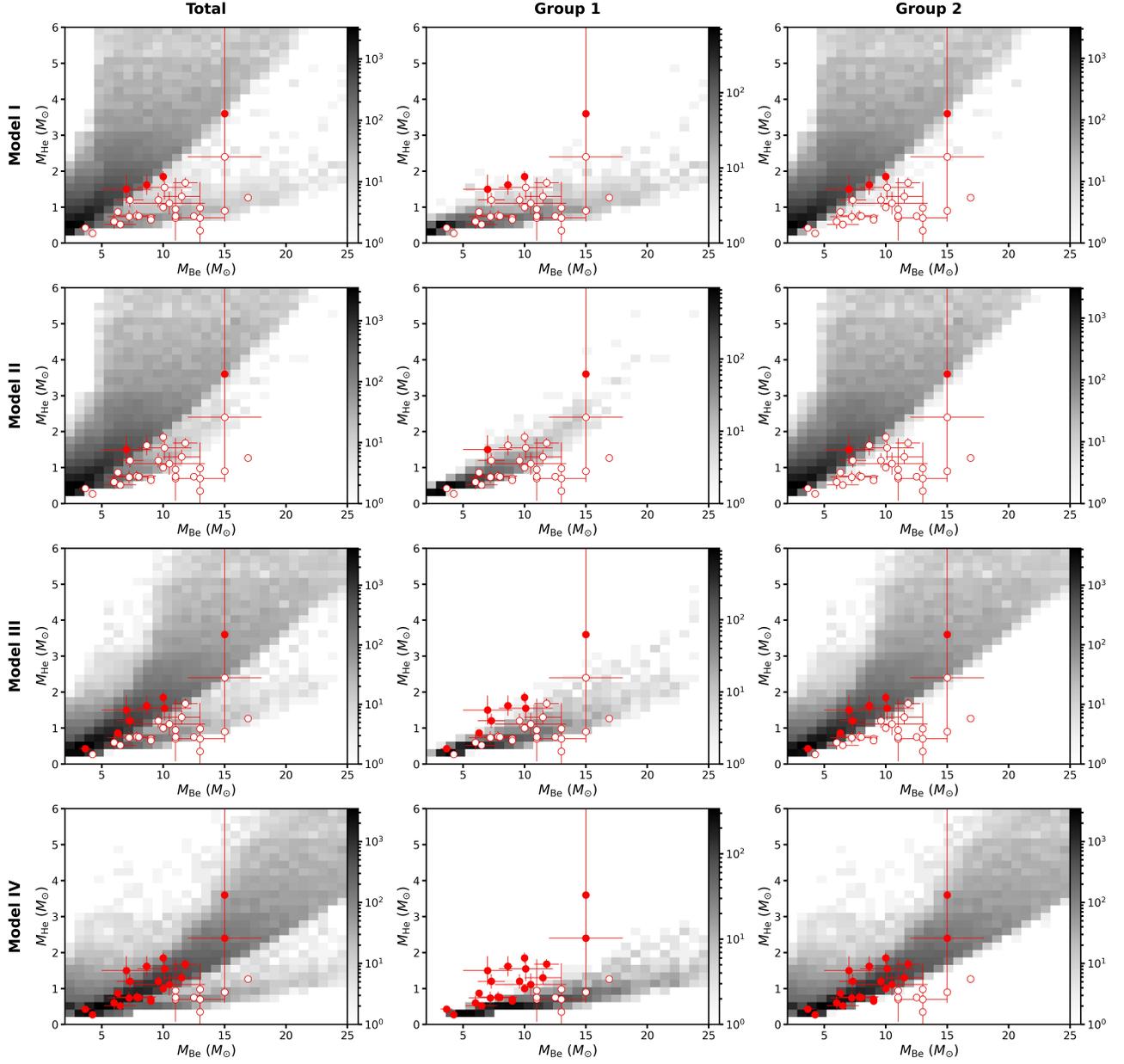

**Figure 2.** The simulated distributions of Galactic BeHe binaries in the $M_{\rm Be}$–$M_{\rm He}$ plane for models I–IV (from top to bottom). The simulations are divided into two groups based on different Be star evolutionary paths, as shown in the middle (group 1) and right (group 2) panels. Each panel includes a 30 × 30 matrix element for the corresponding binary parameters. The color of each pixel represents the number of BeHe binaries in the corresponding matrix. The observed BeHe binaries are classified into group 1 or group 2 based on the corresponding population synthesis results. We compare the numbers in the pixels of the middle panels and right panels, where the observed samples (the most probable values) are located. The observed binaries are assigned to group 1 if the corresponding numbers are larger or equal, as shown by red open circles; otherwise, they are assigned to group 2, as shown by red filled circles.

The distributions of $M_{\rm Be}$ and $M_{\rm He}$ are shown in Figure 2, where the observations are also shown for comparison. Following the results discussed in Section 3.1, we divide the simulated results and the observations into two groups, as shown in the middle and right panels, respectively. For the observed BeHe binaries, it is hard to know the specific mass transfer history; thus, the classification cannot be performed directly. Therefore, we classify the observed sample according to the population synthesis results of Figure 2. Specifically, we compare the corresponding numbers in the pixels of the middle panels (group 1) and right panels (group 2), where the observational samples (the most probable values) are located. The observational sample will be classified into group 1 if the corresponding numbers are larger (or equal); otherwise, it will be classified into group 2. In model I and model II, only four and two observed Be star binaries with relatively low mass ratios fit into group 2, respectively. There are more Be star binaries with large mass ratios due to the high mass transfer efficiencies, and more observed Be star binaries will fit into group 2 with the increase of mass transfer efficiency. The fraction of observed Be star binaries in group 2 is one-third and two-thirds for model III and model IV, respectively.

In Figure 3, we present the number distributions in the $M_{\rm Be}$–$P_{\rm orb}$ diagram. In model I, though the mass transfer efficiency varies in a large range, the mass transfer of most binaries ($\gtrsim 95\%$) in group 2 is very inefficient ($\beta \lesssim 0.1$). Then the

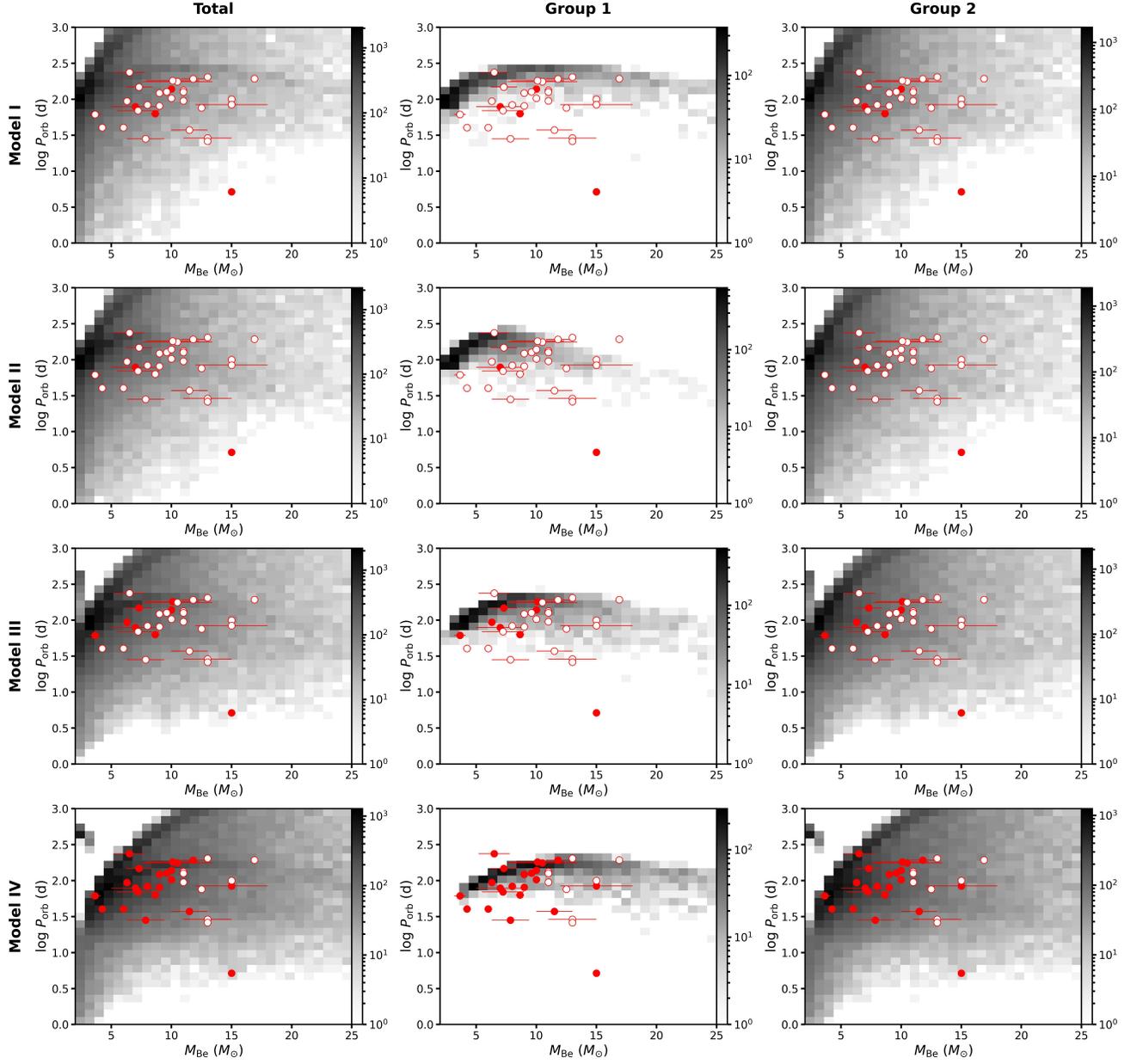

**Figure 3.** Same as Figure 2 but for the simulated distributions of Galactic BeHe binaries in the $M_{Be}$–$P_{orb}$ plane for models I–IV (from top to bottom).

number density distributions of the orbital periods in model I and model II are quite similar. From model II to model IV, the mass transfer efficiency increases, and the number density peak of the orbital period distribution (the darkest regions) moves to a lower value, which is consistent with that of Y. Shao & X. D. Li (2021). According to the results in Figure 2, the observed Be star binaries are also distinguished into two groups. In model I, most observed Be star binaries are fitted into group 1. However, the simulations in group 1 predict typically higher orbital periods than those of the observations. From model II to model IV, the number of observed Be star binaries in group 2 increases with increasing mass transfer efficiencies. As a whole, models III and IV support the observations better. In combination with the results in Figures 2 and 3, our simulations suggest that the formation of Be stars may need a high mass transfer efficiency ($\gtrsim 0.5$) to support the observations better, consistent with many previous works

(S. Vinciguerra et al. 2020; T. Lechien et al. 2025; M. Nuijten & G. Nelemans 2025; C. Schürmann et al. 2025; X. T. Xu et al. 2025). Notably, observations reveal an apparent scarcity of He stars with masses $\gtrsim 2\,M_\odot$, a phenomenon that may be attributed to significant selection effects (e.g., Y. Götberg et al. 2018; L. Yungelson et al. 2024; B. Hovis–Afflerbach et al. 2025).

Interestingly, there are BeHe binaries with Be star masses in the range of 2–4 $M_\odot$ and orbital periods around 400 days in model III and model IV (the upper left region in the left panels of model III and model IV). These binaries are not found in Y. Shao & X. D. Li (2021). The origin of them can be understood as follows. This part of BeHe binaries is produced from the stable mass transfer of low-mass RGB stars ($\sim$1–2 $M_\odot$). In our simulations, we adopt the mass transfer stability criterion of H. Ge et al. (2010, 2015, 2020a, 2024), and the critical mass ratios for donors at the RGB stage are typically in the range of 1.2–1.5. In contrast, in Y. Shao &

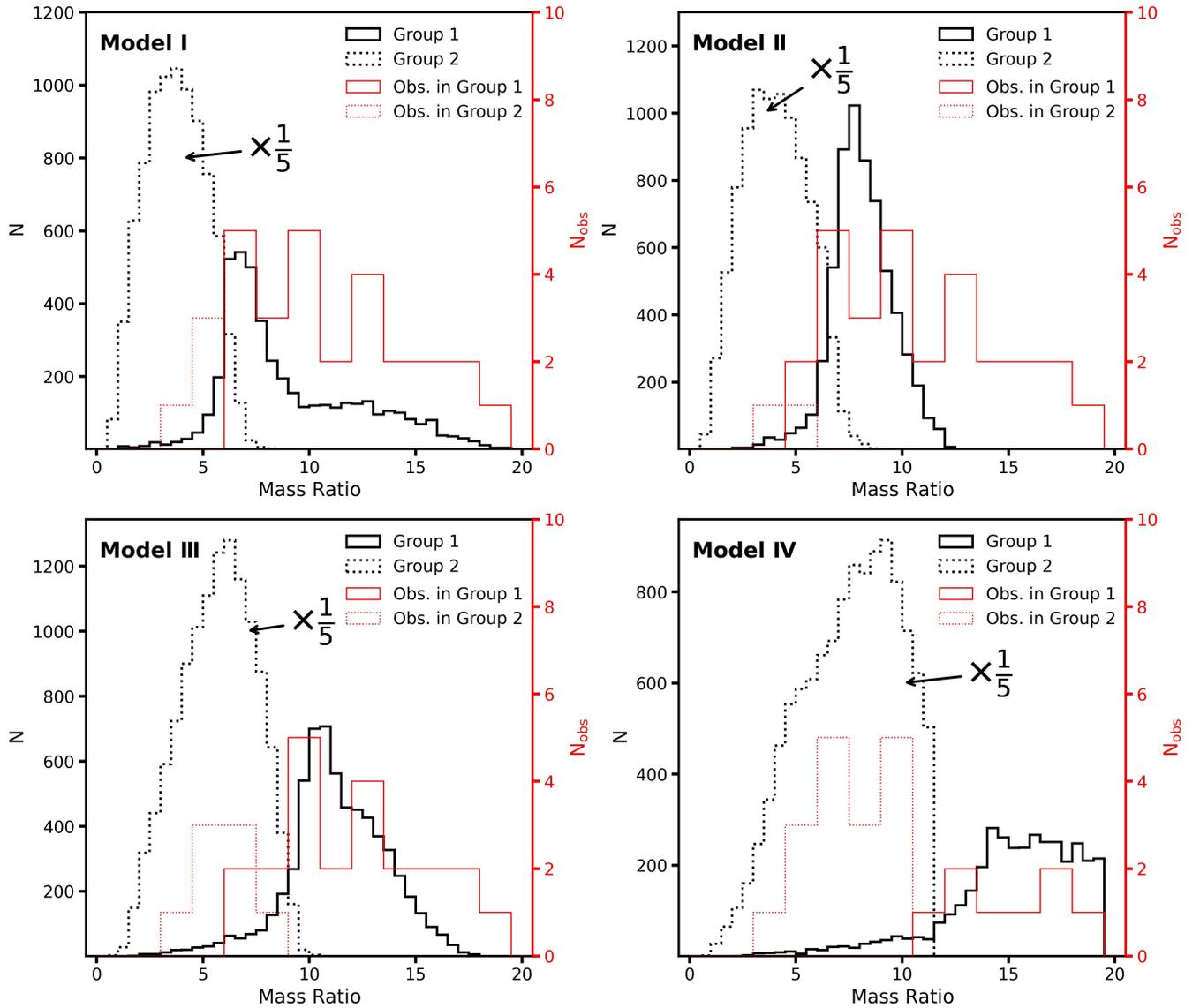

**Figure 4.** The number distributions of Galactic BeHe binaries as a function of the mass ratio ($M_{Be}/M_{He}$) in models I–IV. The black solid and dotted lines are for BeHe binaries in group 1 and group 2, respectively, where the numbers of group 2 have been decreased by 5. The red solid curves correspond to observational BeHe binaries in group 1 (Obs. in Group 1), and the red dotted curves correspond to observational BeHe binaries in group 2 (Obs. in Group 2).

X.-D. Li (2014), the mass transfer is always unstable when the donor has climbed to the giant branch. It should be noted that we only focus on the Be star with an He star companion, and the He core can only be ignited as the mass transfer starts for the donor close to the tip of the RGB (Z. Han et al. 2002, 2003; X. Chen & Z. Han 2003; X. Chen et al. 2013). Therefore, only primordial binaries with orbital periods in a very narrow range can produce BeHe binaries. These BeHe binaries are also not shown in model I and model II. The reason is that in both models, the mass transfer is generally very inefficient and finally produces Be stars with mass less than $2\,M_\odot$, which is not shown in our simulations.

In the next step, we make a quantitative analysis of the results in group 1 and group 2. The mass ratio distributions are shown in Figure 4, where the numbers of group 2 have been decreased by 5 for clarity. In model I and model II, we see that the shapes of the mass ratio distributions of group 2 are quite similar, since the mass transfer of most binaries in group 2 of model I is quite inefficient. For binaries in group 1, the mass ratio distribution of model I has a peak around 7 with an extended trail to 20, while the peak in model II is around 8 with a maximum mass ratio value of 12.5. The reason for the large dispersion in model I is due to the varied mass transfer efficiencies in the range of 0.07–0.80. In model III and model IV, the mass transfer efficiencies are typically higher. Then, we see that the peaks for group 1 and group 2 shift toward higher values. It is interesting to see that the observed distributions of Be star binaries have peaks similar to those of the results in model III and model IV. However, we realize that the numbers of group 2 are much higher than that of the observations. There are two potential reasons. First, this may be due to strong selection effects, which have prevented many low mass ratio Be star binaries from being discovered. Second, many binaries going through case B mass transfer may not produce Be stars due to some unknown physical factors. This finding is similar to that of the recent work of M. Nuijten &

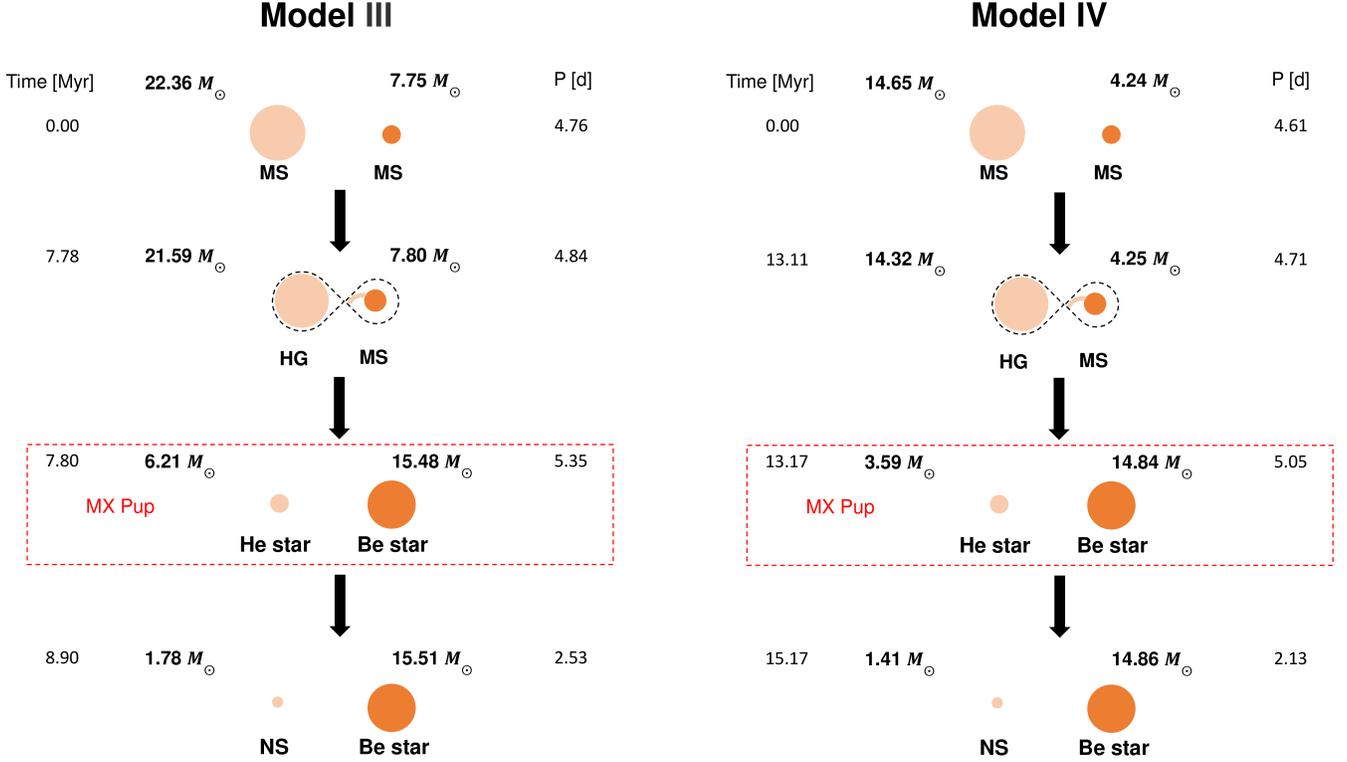

**Figure 5.** The typical evolutionary history of MX Pup in model III ($\beta = 0.5$) and model IV ($\beta = 1$). The mass transfer phases both happen at the HG phase. The observed parameters of MX Pup are $M_{Be} \approx 15\,M_\odot$, $M_{He} \approx 0.6$–$6.6\,M_\odot$, and $P_{orb} = 5.1526 \pm 0.0011$ days.

G. Nelemans (2025), who studied the formation of WR + O binaries and found that most observed systems experienced case A mass transfer instead of case B. The physical explanations of this phenomenon need to be investigated in more detail.

### 3.3. Formation Scenario of MX Pup

MX Pup is an exciting Be star binary with an orbital period of only 5.15 days (R. E. Mennickent et al. 1994, 1997; F. Carrier et al. 2002; Y. R. Cochetti et al. 2019, 2023; Y. Nazé et al. 2022; L. Wang et al. 2023) that can help to contain the theoretical models. In our simulations, models III and IV can successfully reproduce the observed parameters of MX Pup. However, in the work of Y. Shao & X. D. Li (2021), their models with similar inputs cannot produce the Be star binary in such a close orbit (their models II and III). In Figure 5, we present the typical evolutionary history of MX Pup in model III and model IV. We see that in both models, the initial mass ratios are well above 2.5, and the mass transfer phases happen at the HG stage. Binaries with the same initial parameters would enter into the CE phase and do not produce Be star binaries in Y. Shao & X. D. Li (2021), as their critical mass ratios are typically less than 2 in their models II and III. The mass transfer from the massive one to the less massive one generally leads to orbital shrinkage (T. M. Tauris & E. P. J. van den Heuvel 2006), and then binaries with large mass ratios would more likely produce binaries with short orbits.

The largest difference between model III and model IV is the mass transfer efficiency. Model III has a relatively lower mass transfer efficiency, and then the primordial binary has large masses to obtain the Be star with a mass close to the observation. As a result, the produced He star in model III also has a large mass of $6.21\,M_\odot$, which is higher than that in model IV (with $3.59\,M_\odot$). In the observations, the He star mass in MX Pup is very uncertain. Therefore, our results suggest that the precise estimation of the He star mass can help to constrain the mass transfer efficiency.

### 4. Conclusion

This study aims to understand the stable RLOF processes by constructing BeHe binary populations with an updated mass transfer stability criterion. Our main conclusions are summarized as follows.

(1) The theoretical simulations show that the BeHe binary populations can be divided into two groups with distinct physical properties. In group 1, the mass transfer happens at the early-MS stage, and in group 2, the mass transfer happens at a more evolved stage. We found that BeHe binaries in group 1 typically have a larger mass ratio ($M_{Be}/M_{He}$) than those in group 2. The results show an approximately negative correlation between the mass ratios of BeHe binaries and the initial mass ratio of the primordial binaries in group 2. We may put a constraint on the critical mass ratio with this feature.

(2) The observational samples of Be star binaries are also divided into two groups according to the theoretical $M_{Be}$–$M_{He}$ distributions (see Figure 2). For each model, we compare the predicted numbers in the pixels of the two groups where the observed samples are located. The observed binaries are assigned to group 1 if the corresponding numbers are larger or equal; otherwise, they are assigned to group 2. While this classification method is somewhat arbitrary, it enables a quantitative comparison between the simulations and observations. Our BPS simulations suggest that a higher mass transfer efficiency ($\gtrsim 0.5$) supports the observations better. In particular, in models III and IV, the peak values of the mass ratios are consistent with the observations. However, we found that the simulations predicted too many BeHe binaries in group 2, which disagrees with the observations. One possible reason is that there are still many low mass ratio Be star binaries that have not been discovered. On the other hand, binaries going through case B mass transfer may not produce Be stars due to some unknown physical factors.

(3) Models III and IV can also help to explain a special BeHe binary, MX Pup, with an orbital period of 5.15 days. To explain the formation of this binary, the initial mass ratio needs to be larger than 2.5, so that the orbit would not expand too much during the mass transfer phase. The precise estimation of the He star mass for MX Pup may put a constraint on the mass transfer efficiency.

## Acknowledgments

The authors thank the anonymous referee who has provided constructive and helpful comments to improve the paper. This work is supported by the Natural Science Foundation of China (grant Nos. 12288102, 12125303, 12090040/3, 12473034, 12273105, 11703081, 11422324, 12073070, 12173081), the Strategic Priority Research Program of the Chinese Academy of Sciences (grant Nos. XDB1160201, XDB1160000), the National Key R&D Program of China (grant Nos. 2021YFA1600403, 2021YFA1600400), the Yunnan Revitalization Talent Support Program-Science & Technology Champion Project (No. 202305AB350003), the International Centre of Supernovae (ICESUN), Yunnan Key Laboratory of Supernova Research (Nos. 202302AN360001, 202201BC070003), Yunnan Fundamental Research Projects (No. 202401AT070139), the Key Research Program of Frontier Sciences of CAS (No. ZDBS-LY-7005).

## ORCID iDs

Yuchen Bao https://orcid.org/0009-0006-5929-4199
Zhenwei Li https://orcid.org/0000-0002-1421-4427
Hongwei Ge https://orcid.org/0000-0002-6398-0195
Xuefei Chen https://orcid.org/0000-0001-5284-8001
Zhanwen Han https://orcid.org/0000-0001-9204-7778